# Ultrafast dynamics of light-induced spin crossovers under high pressure

Yu.S.Orlov[1,2], S.V.Nikolaev[1,2], A.I.Nesterov[3], S.G.Ovchinnikov[1,2]

[1]Kirensky Institute of Physics, Federal Research Center KSC SB RAS, Krasnoyarsk, 660036 Russia
[2]Siberian Federal University, Krasnoyarsk, 660041 Russia
[3]CUCEI, Universidad de Guadalajara, Guadalajara, CP 44420, Jalisco, Mexico

Within the multielectron model of magnetic insulator with spin crossover under high pressure we have studied the mean field phase diagrams in pressure-temperature plane and dynamics of a sudden excited non equilibrium spin state. We obtain the different relaxation of the magnetization, high spin/low spin occupation numbers, and the metal-oxygen bond length for different values of the external pressure. We found the long living oscillations of magnetization without pressure and at small external pressure. Close to crossover pressure the smooth relaxation is accompanied with a set of sharp strongly non linear oscillations of magnetization and HS/LS occupation numbers that are accompanied by the Franck-Condon resonances.

1. **Introduction**

The ultrafast magnetism is a very active area in modern condensed matter physics [1-7]. With the femtosecond pump – probe technique a lot of exciting results have been obtained for different magnetic materials including metals and insulators, among them the ultrafast demagnetization or long living magnetization precession. We will discuss here the ultrafast magnetic dynamics in materials with spin crossover, when switching between the high spin (HS) state and the low spin (LS) state is induced by some external impact like high pressure (typically iron oxides) or temperature (typically metal-ligand complexes in organic matrix) [8-10]. The HS-LS transition has been found also under light irradiation and called the LIESST effect (Light Induced Spin State Trapping) [8,9]. The LIESST effect in $Fe(phen)_2(NCS)_2$ has been studied recently by time-resolved XANES and optical spectroscopy at the XPP LCLS XFEL (The X-ray Pump-Probe instrument at the Linac Coherent Light Source) in Stanford [11]. The light-induced LS-HS switching and the forthcoming relaxation has revealed local deformation and vibronic oscillations of ligands.

The other group of materials where HS-LS transitions are induced by high pressure is the Fe-based oxides with $Fe^{3+}$ or $Fe^{2+}$ ions [12-17] with the HS ground state and spin crossover at Pc close to 50-60 GPa. These oxides are typical Mott insulators with electronic structure and properties determined by strong electronic correlations [18,19]. There is one more unique group of $3d$ - oxides with the LS ground state that demonstrates spin crossover with heating, the rare-earth cobaltites $LnCoO_3$. $LaCoO_3$ is one paradigmatic example where the strong electron, spin, and lattice coupling induced by electronic correlations results in a low-temperature spin transition and a high-temperature semiconductor-to-metal transition that is still not completely understood. Recently, the ultrafast metallization in $LaCoO_3$ using time-resolved soft x-ray reflectivity experiments has been revealed [20]. Metallization is shown to occur via transient electronic, spin, and lattice separation.

A simple picture of spin crossover is based on the single $3d$ ion in the ligand crystal field picture and conceptually is quite simple. The intraatomic Coulomb interaction results in the formation of the HS electronic configuration (the Hund rule) with the Hund exchange energy $J_H$ gain. Nevertheless in the crystal the large value of the cubic crystal field $10Dq$ may stabilize the LS state. There is a competition between the Hund exchange and the crystal field. From the Tanabe-Sugano diagrams it is clear that spin crossover may occur for $d^n$ ionic compounds with $n = 4 - 7$ [21]. Within the single site model the spin crossover at $T = 0$ is a quantum phase transition with the Berry-type phase being the order parameter [22]. The simple single site model cannot answer to several important questions. Is the spin crossover a phase transition or not? What are the effects of cooperativity that may be induced by the interatomic exchange interactions or by interaction with lattice?

In the literature there are several simplified models discussing effects of cooperativity and the influence of pressure, temperature and irradiation on spin crossovers [23-32]. The vibron model of the metal-ligand complexes [33] incorporates the spin crossover and the change of the local vibrations in the metal-ligand complexes within non adiabatic theory of electron-vibron interaction. Molecular dynamics calculations using the stochastic Monte-Carlo approach [34, 35] allows to describe the photoinduced transition beyond the Born-Oppenheimer approximation [36]. For magnetic oxides both mechanisms of cooperativity are important: the interatomic exchange interaction and cooperativity of magnetic cations via electron-vibron-electron interaction. In the present paper we will study the effect of both cooperativity mechanisms on spin crossover.

We consider the multielectron model of magnetic oxide with two local $d^n$ terms (HS and LS) with interatomic exchange between cations. The electron-vibron interaction in this problem is especially important due to the large (about 10%) difference of the HS and LS ionic radii, so the transition of the HS into LS state and back results in a strong Me-O bond length contraction-dilatation. The local excitations between the HS and LS terms that results from the spin-orbit interaction and mix the HS and LS states are also important and will be considered. Within this model we study both the equilibrium thermodynamics and non equilibrium system dynamics.

The paper is organized as follows. In Sec.II we describe the model and derive a system of the mean field equations for magnetization $m$, HS concentration $n$ and the Me-O bond length $q$. The mean field phase diagrams (pressure-temperature) are discussed in Sec.III. Sec.IV contains the application of the master equation for our dynamical quantities within the Redfield approximation. Numerical results for system dynamics at various pressures are analyzed in Sec.V. Discussion of the results is given in Sec.VI.

2. **The effective Hamiltonian of the magnetic insulator with spin crossover**

We consider a 3D lattice with the $3d^n$ ions at every site surrounded by $z$ ligands with the equilibrium Me-O bond length $l_0$. Instead of full set of multielectron terms we consider only two of them, the HS and LS with the energies $E_{HS}$ and $E_{LS}$, these energies are equal at some pressure $P_{C0}$ according to a single site model. To describe the interatomic exchange interaction we use the Hubbard $X$ - operators constructed with the help of all HS and LS states. It can be done for arbitrary values of the HS and LS, nevertheless we specify the $S_{HS} = S = 2$ and $S_{LS} = 0$

. This choice corresponds to 3 $d^6$ ions (FeO и Mg$_{1-x}$Fe$_x$O). Then the complete and orthogonal local set of electronic eigenstates includes the HS quintet with spin projection $|\sigma\rangle$, $\sigma = -S, -S+1, ... + S$ and LS- singlet $|s\rangle$.

We write down the model Hamiltonian in the following way

$$\hat{H}_{eff} = \hat{H}^{(S)} + \hat{H}^{(e,q)} + \hat{H}^{(ex)}. \quad (1)$$

Here the first term (see Eq.(2)) describes magnetic cations with the interatomic exchange interaction $J$ beyond the conventional Heisenberg model due to the presence of the LS term and provides the magnetic cooperativity. The LS term does not allow to write down the magnetization via the Brillouin function. The analysis of the non Heisenberg effects in such model has been discussed recently in the Ref. [37].

$$\hat{H}^{(S)} = J \sum_{\langle i,j \rangle} \left( \hat{\vec{S}}_i \hat{\vec{S}}_j - \frac{1}{4} \hat{n}_i \hat{n}_j \right) + E_{LS} \sum_i X_i^{s,s} + E_{HS} \sum_{i,\sigma=-S}^{+S} X_i^{\sigma,\sigma}. \quad (2)$$

The two last terms in Eq.(2) describe the LS and HS states separated by a spin gap $\Delta_S = E_{LS} - E_{HS}$, that linearly decreases with pressure and changes its sign at the $P = P_{C0}$, the crossover pressure in the single site approach. The spin operator $\hat{\vec{S}}_i$ for $S = 2$ is written in the Hubbard operator representation [38, 39]

$$\hat{S}_i^+ = 2X_i^{-1,-2} + \sqrt{6}X_i^{0,-1} + \sqrt{6}X_i^{+1,0} + 2X_i^{+2,+1}, \quad \hat{S}_i^- = 2X_i^{-2,-1} + \sqrt{6}X_i^{-1,0} + \sqrt{6}X_i^{0,+1} + 2X_i^{+1,+2},$$

$$\hat{S}_i^z = -2X_i^{-2,-2} - X_i^{-1,-1} + X_i^{+1,+1} + 2X_i^{+2,+2};$$

$\hat{n}_i = 6 \sum_{\sigma=-S}^{+S} X_i^{\sigma,\sigma} + 6 X_i^{s,s}$ is the operator of electron number at the site $i$, $\langle \hat{n}_i \rangle = 6$. The condition of completeness of the HS and LS set of states yields the sum rule $\sum_{\sigma=-S}^{+S} X^{\sigma,\sigma} + X^{s,s} = 1$.

The effective Hamiltonian (1) has been obtained [40] from the miscoscopic multiband p-d model using the projection Hubbard operators within the multielectron approach LDA+GTB to the electronic properties of strongly correlated materials [41,42]. The second term (see Eq. (3)) in the Hamiltonian describes the energy of ligand octahedral intramolecular full symmetrical vibrations, the electron-vibron interactions [43, 44] and the elastic interatomic interaction. This term provides the cooperativity throw the elastic lattice and is responsible for the volume change under external pressure and temperature. It is given by

$$\hat{H}^{(e,q)} = \sum_i \left( \frac{1}{2} k \hat{q}_i^2 + \frac{\hat{p}_i^2}{2M} \right) - \sum_i \left( g_1 \hat{q}_i + g_2 \hat{q}_i^2 \right) \left( -X_i^{s,s} + \sum_{\sigma=-S}^{+S} X_i^{\sigma,\sigma} \right) - \frac{1}{2} V_q \sum_{\langle i,j \rangle} \hat{q}_i \hat{q}_j, \quad (3)$$

where $g_1$ и $g_2$ are the parameters of linear and quadratic electroт-vibron interaction within the MeO$_6$ octahedra, $k$ is the elastic parameter, $M$ is the anion mass, $\hat{q}_i$ is the normal coordinate operator of the Me-O breathing vibration, and $\hat{p}_i$ is corresponding momentum operator, $V_q$ is the parameter of interatomic eleastic coupling. The Me-O bond length is equal to $l = l_0 + \langle \hat{q} \rangle$, where $l_0$ is the equilibrium bond length. Due to the large difference in the HS and LS ionic radii we have to include in (3) the anharmonic coupling $g_2$. It results in the renormalization of the elastic parameter, for the HS $k_{HS} = k - 2g_2$ and for the LS $k_{LS} = k + 2g_2$. The third contribution (4) to the Hamiltonian describes the excitations between the HS and LS terms induced by the spin-orbital interaction [45] that resulted to mixing of the HS and LS states. It can be written as

$$\hat{H}^{(ex)} = J_x \sum_i \sum_{\sigma=-S}^{+S} \left( X_i^{s,\sigma} + X_i^{\sigma,s} \right) \qquad (4)$$

In general the Hamiltonian (1) describes very complicated many body physics with interacting spin, charge and lattice degrees of freedom. That is why we will treat it in the mean field approximation for both interatomic interactions, the exchange $J$ in Eq.(2) and the elastic term $V_q$ in Eq.(3). In the mean field approximation the Hamiltonian is given by

$$\hat{H}_{MF} = H_0 - B \sum_i \hat{S}_i^z + \Delta_S \sum_i X_i^{s,s}$$
$$+ \sum_i \left( \frac{1}{2} k \hat{q}_i^2 + \frac{\hat{p}_i^2}{2M} \right) - \sum_i \left( g_1 \hat{q}_i + g_2 \hat{q}_i^2 \right) \left( -X_i^{s,s} + \sum_{\sigma=-S}^{+S} X_i^{\sigma,\sigma} \right) - V_q \langle \hat{q} \rangle \sum_i \hat{q}_i + \qquad (5)$$
$$+ J_x \sum_i \sum_{\sigma=-S}^{+S} \left( X_i^{s,\sigma} + X_i^{\sigma,s} \right)$$

Here $B = zJSm$ is the molecular Weiss field, where $z = 6$ is the number of nearest neighbors, $m = \dfrac{\langle \hat{S}^z \rangle}{S}$ is the normalized magnetization of sublattice in a two sublattice antiferromagnet.

To write down the matrix of the mean-field Hamiltonian (5) we choose the local basis functions as a product of spin and harmonic oscillator eigenfunctions. The spin eigenfunctions look like

$$|\alpha, s_z\rangle, \quad s_z = -S, (-S+1), \ldots, +S \text{ for HS state } (\alpha = 1) \text{ and } s_z = 0 \text{ for LS state } (\alpha = 2).$$

For harmonic oscillator we introduce phonon creation and annihilation operator as usually by relations $\hat{q}_i = \sqrt{\dfrac{1}{2M\omega}} \left( a_i + a_i^\dagger \right)$ and $\hat{p}_i = \dfrac{1}{i}\sqrt{\dfrac{M\omega}{2}} \left( a_i - a_i^\dagger \right)$, so the eigenstate with $n_{ph} = 0, 1, 2, \ldots$ is given by $|n_{ph}\rangle = \dfrac{1}{\sqrt{n_{ph}!}} \left( a^\dagger \right)^{n_{ph}} |0, 0, \ldots, 0\rangle$. Finally, our basis functions are given by $|\alpha, s_z, n_{ph}\rangle = |\alpha, s_z\rangle |n_{ph}\rangle$.

In this basis the matrix of Hamiltonian (5) can be written as

$$H^{\alpha's_z'n_{ph}'}_{\alpha s_z n_{ph}} = \left\{\left[-\frac{\Delta_S}{2} - \frac{g_2 \hbar\omega}{2k}(2n_{ph}+1)\right]\lambda_\alpha + \hbar\omega\left(n_{ph}+\frac{1}{2}\right) - (zJSm)s_z + H_0/N\right\}\delta_{s_z s_z'}\delta_{\lambda_\alpha \lambda_{\alpha'}}\delta_{n_{ph} n_{ph}'} +$$

$$+ J_x \delta_{\lambda_\alpha, -\lambda_\alpha'}\delta_{n_{ph} n_{ph}'} -$$

$$-\lambda_\alpha g_1 \sqrt{\frac{\hbar\omega}{2k}}\left(\sqrt{n_{ph}}\,\delta_{n_{ph}-1,n_{ph}'} + \sqrt{n_{ph}+1}\,\delta_{n_{ph}+1,n_{ph}'}\right)\delta_{\lambda_\alpha \lambda_{\alpha'}}\delta_{s_z s_z'} -$$

$$-V_q\langle\hat{q}\rangle\sqrt{\frac{\hbar\omega}{2k}}\left(\sqrt{n_{ph}}\,\delta_{n_{ph}-1,n_{ph}'} + \sqrt{n_{ph}+1}\,\delta_{n_{ph}+1,n_{ph}'}\right)\delta_{\lambda_\alpha \lambda_{\alpha'}}\delta_{s_z s_z'} -$$

$$-\lambda_\alpha g_2 \frac{\hbar\omega}{2k}\left(\sqrt{n_{ph}(n_{ph}-1)}\,\delta_{n_{ph}-2,n_{ph}'} + \sqrt{(n_{ph}+2)(n_{ph}+1)}\,\delta_{n_{ph}+2,n_{ph}'}\right)\delta_{\lambda_\alpha \lambda_{\alpha'}}\delta_{s_z s_z'} \tag{6}$$

where $N$ is the number of lattice sites; $\lambda_\alpha = 1$, for $\alpha = 1$ и $\lambda_\alpha = -1$, for $\alpha = 2$.

The eigenfunctions of the Hamiltonian (5) can be written in polaronic representation

$$|\varphi_k\rangle = \sum_{n_{ph}=0}^{N_{ph}}\left[a_{n_{ph},k}|2,0,n_{ph}\rangle + \sum_{s_z=-S}^{+S} b_{n_{ph},s_z,k}|1,s_z,n_{ph}\rangle\right]. \tag{7}$$

Without electron-vibron interaction the electronic part of this eigenfunction describes a superposition of the LS singlet and the HS term with (2$S$+1) spin projections, in our case with the $S=2$ configuration Fe$^{2+}$ with $t_{2g}^4 e_g^2$ has also 3-fold orbital degeneracy that is not shown in Eq.(7) for simplicity. Nevertheless it is included in the numerical calculations by a factor 3. So the number of the HS sublevels is 15 and the total number of electronic sublevel at each site is $N_e = 16$. The polaronic representation (7) treat the local electron-vibron interaction exactly and describes the superposition of the $N_e$ states without vibron, with one vibron, etc. The cut-off vibron number $N_{ph}$ is found from the condition that addition one extra vibron changes the ground state $|\varphi_0\rangle$ energy less than 1%, $E_0(N_{ph}+1) \approx E_0(N_{ph})$ and coefficients $a_{n_{ph},0}(N_{ph}+1) \approx a_{n_{ph},0}(N_{ph})$, $b_{n_{ph},s_z,0}(N_{ph}+1) \approx b_{n_{ph},s_z,0}(N_{ph})$ (our computation error is less than 1%) (For finite temperature computations we also have checked similar properties for several excited eigenstates $|\varphi_k\rangle$ both for the energy $E_k$ and coefficients $a_{n_{ph},k}(N_{ph}+1) \approx a_{n_{ph},k}(N_{ph})$, $b_{n_{ph},s_z,k}(N_{ph}+1) \approx b_{n_{ph},s_z,k}(N_{ph})$). In other words, $N_{ph}$ determines the number of vibrons that has to be included for the given set of parameters to form the vibron cloud around electron in the ground and several excited polaronic states. In our computations $N_{ph} = 300 \div 500$ depending on the model parameters, temperature and pressure. The multivibron contribution to the eigenstates (7) results in the Franck-Condon resonances during their excitations [46].

With the eigenfunctions (7) one can obtain the quantum mechanical averages of the HS concentration $\hat{n}$, bond length deformation $\hat{q}$ and sublattice magnetization $\hat{S}^z$

$$\langle\hat{n}\rangle_k = \left\langle\varphi_k\left|\sum_\sigma X^{\sigma,\sigma}\right|\varphi_k\right\rangle = \sum_{n_{ph}=0}^{N_{ph}}\sum_{s_z=-S}^{+S}\left|b_{n_{ph},s_z,k}\right|^2, \tag{8}$$

$$\langle \hat{q} \rangle_k = \langle \varphi_k | \hat{q}_k | \varphi_k \rangle = \sqrt{\frac{\hbar}{2M\omega}} \sum_{n_{ph}=0}^{N_{ph}} \left\{ \sqrt{n_{ph}} \left( a_{n_{ph},k} a_{n_{ph}-1,k} + \sum_{s_z=-S}^{+S} b_{n_{ph},s_z,k} b_{n_{ph}-1,s_z,k} \right) + \sqrt{n_{ph}+1} \left( a_{n_{ph},k} a_{n_{ph}+1,k} + \sum_{s_z=-S}^{+S} b_{n_{ph},s_z,k} b_{n_{ph}+1,s_z,k} \right) \right\}, \quad (9)$$

$$\langle \hat{S}^z \rangle_k = \langle \varphi_k | \hat{S}^z | \varphi_k \rangle = \sum_{n_{ph}=0}^{N_{ph}} \sum_{s_z=-S}^{+S} s_z |b_{n_{ph},s_z,k}|^2. \quad (10)$$

After thermodynamic averaging we get a system of self consistent equations (11-13)

$$n = \langle \hat{n} \rangle = \sum_k \frac{\langle \hat{n} \rangle_k e^{-E_k/k_B T}}{Z}, \quad (11)$$

$$q = \langle \hat{q} \rangle = \sum_k \frac{\langle \hat{q} \rangle_k e^{-E_k/k_B T}}{Z}, \quad (12)$$

$$m = \frac{\langle \hat{S}^z \rangle}{S} = \frac{1}{S} \sum_k \frac{\langle \hat{S}^z \rangle_k e^{-E_k/k_B T}}{Z}, \quad (13)$$

where $Z = \sum_k e^{-E_k/k_B T}$.

Before turning to numerical simulations, we would like to discuss typical for $3d$ - oxides parameters. The most studied at high pressure are, $Fe_2O_3$ and some other oxides with $3d^5$ $Fe^{3+}$ ion that has HS value $S = 5/2$ and LS $S = 1/2$ with $P_C = 47$ GPa for $FeBO_3$ [16]. We consider in this paper spin crossover in oxides with $3d^6$ ions that have HS $S = 2$ and LS $S = 0$, the example is given by $Fe_xMg_{1-x}O$ with $P_C = 55$ GPa [47]. The spin gap values for all $3d^n$ ions are given in [48]. The spin gap for $Fe^{2+}$ is equal to $\Delta_S = 2(2J_H - 10Dq)$, where $J_H$ is the intraatomic Hund exchange coupling stabilizing the HS state, and $10Dq$ is the cubic crystal field parameter, stabilizing the LS state. With increasing pressure and decreasing the interatomic distance the crystal field and the effective interatomic exchange interaction linearly grows as $10Dq(P) = 10Dq(0) + \alpha_\Delta P$ [47] and $J(P) = J_0 + bP$ [47]. For Fe ions $J_H = 0.8$ eV and $\Delta_S \sim 1$ eV are typical values. For example, the crystal field at zero pressure $10Dq(0) = 1.57$ eV for $FeBO_3$ has been determined from optical spectra [49-51]. Due to the linear increases crystal field the spin gap can be written as $\Delta_S = a(P_{C0} - P)$ with $a = 2\alpha_\Delta$ and the critical value of pressure $P_{C0}$ that would determine the crossover if there were no cooperativity effects. Due to these effects the critical pressure $P_C$ when the crossover occurs differs from $P_{C0}$ that will be shown in the Sec.III.

Thus, we will take the following model parameters as typical values: $z = 6$, $J_0 = 28$ K, $P_{C0} = 55$ GPa, $a = 80$ K/GPa, $b = 0.5$ K/GPa, $\omega = 0.05$ eV, $k = 7.5$ eV/Å$^2$, $g_1 = 0.8$ eV/Å, $g_2 = 0.75$ eV/Å$^2$, $V_q = 0.2$ эВ/Å.

Due to anharmonic contribution to the electron-vibron interaction (3) the local vibration frequencies are different for HS and LS state $\omega_{HS} = \sqrt{k_{HS}/M}$, $\omega_{LS} = \sqrt{k_{LS}/M}$. For the chosen parameters the frequencies are found to be $\omega_{HS} = 0.045$ eV, $\omega_{LS} = 0.055$ eV. The increasing frequency in the more dense high pressure LS state is evident. The Me-O bond lengths changes are different for the HS and LS states $q_{LS}^0 = -\frac{g_1}{k_{LS}}$, $q_{HS}^0 = \frac{g_1}{k_{HS}}$.

For the chosen set of parameters we obtained $q_{LS}^0 = -0.09$ Å, $q_{HS}^0 = 0.13$ Å, and the difference is equal to $\Delta q^0 = q_{HS}^0 - q_{LS}^0 = 0.22$ Å. At $T = 0$ the bond length $l_0$ is about 2 Å, so $\Delta q^0$ is close to 10% of $l_0$. This difference agrees with typical 10% difference in the LS- and HS- ionic radii. The unit cell volume as a function of pressure and temperature may be written in the following way $V(P,T) = V_r(P,T) + \Delta V(P,T)$, where $V_r(P,T)$ is conventional regular contribution due to the lattice anharmonicity, and additional contribution $\Delta V(P,T) \sim q^3$, due to electron-vibron interaction. Moreover, in materials with spin crossover the redistribution of HS/LS concentrations provides the additional contribution to the lattice dilatation at heating due to large difference of the ionic radii [52].

Finally, via the magnetic anisotropy energy we estimate the mixing of the HS and LS terms value $J_x$, induced by SO interaction. The HS $Fe^{3+}$ term has zero orbital moment and is isotropic. The anisotropy energy induced by the SO interaction appears in the second order of perturbation theory $E_a = J_x^2/\Delta_S$. For HS $Fe^{2+}$ ion the SO interaction in the first order contribution splits the HS term into sublevels with total momentum $\tilde{J} = 1, 2, 3$ but does not mix the HS and LS states [53]. The mixing term (4) appears in the second order of perturbation theory. The typical value of the anisotropy energy $E_a \sim 10K \sim 1$meV. For the spin gap $\Delta_S \sim 1$ eV we get $J_x = 30$ meV. Below we will consider several values of the mixing parameter $J_x$ in the range 10-50 meV.

### 3. The P-T phase diagram

Let us start analysis of the mean field equations (6, 11-13) without the Heisenberg exchange interaction at $J = 0$. Then we obtain $m = 0$ for all pressures and sharp drop of the $n$ and $q$ at the crossover point $P_{C0}$ at $T = 0$. Without cooperativity effect the spin crossover at $T = 0$ is the quantum phase transition with the Berry phase as topological order parameter [22]. It transforms in a smooth crossover for finite temperature. In all phase diagrams we use the rescaled pressure $P/P_{C0}$ and temperature $T/J_0$. Nevertheless even for $J = 0$ we have the other type of cooperativity due to electron-vibron interacton. In the Fig.1 we can see a small but finite temperature range of a sharp crossover close to the critical pressure. For $J = 0$ the crossover from paramagnetic HS to non magnetic LS state is accompanied by isostructural phase transition with the change of volume (Fig. 1b, d).

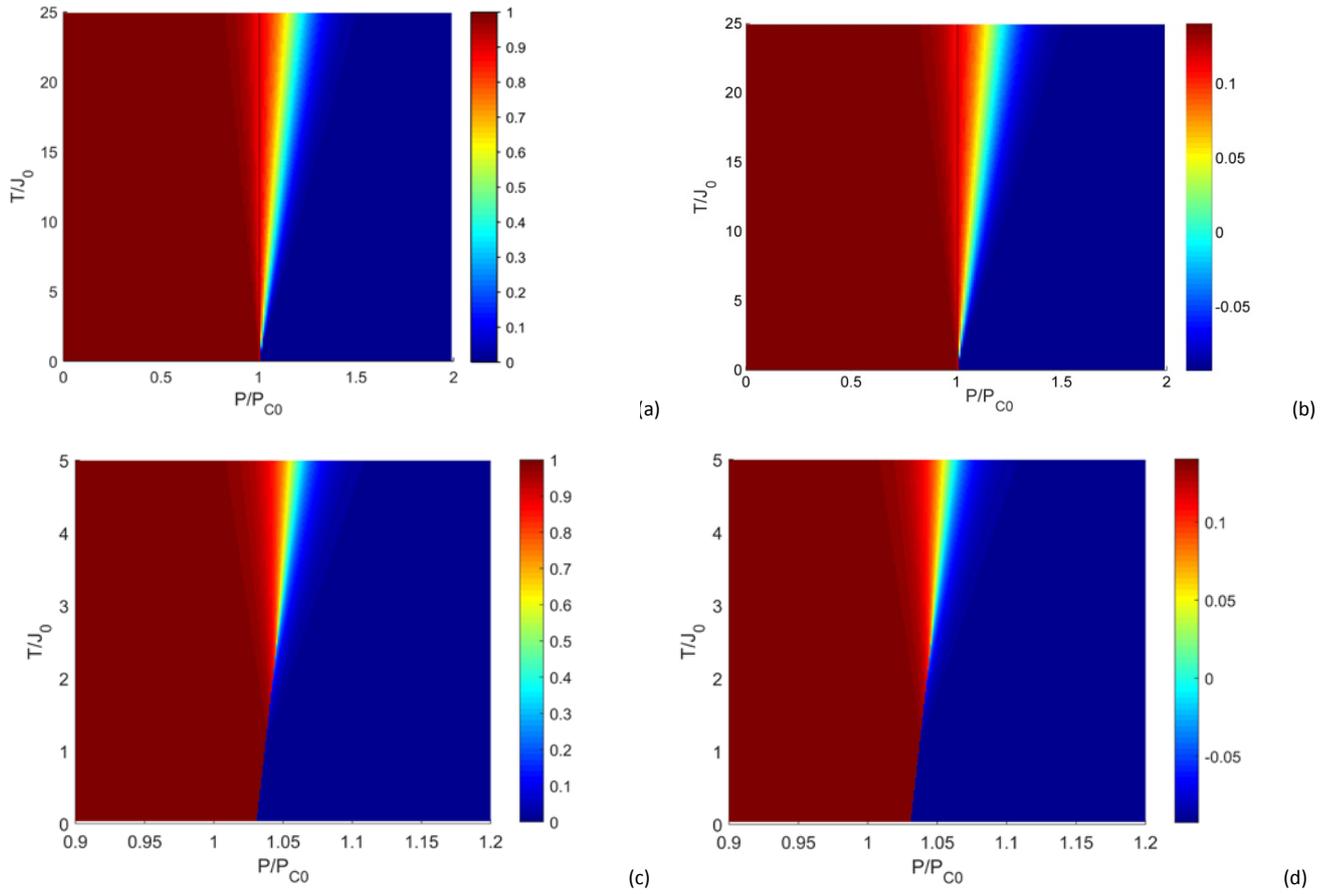

Fig. 1 The $P$-$T$ diagrams of the HS concentration $n$ (a, c) and the bond length deviation $q$ (b, d) when $J = 0$. The (c) and (d) shows the sharp spin crossover at low temperatures close to the critical point.

Fig.2 show maps of all thermodynamic characteristics: HS concentration $n$ (a), magnetization $m$ (b) and displacement $q$ (c) for $J \neq 0$. For some $P$, $T$ values we find several solutions for parameters $n$, $m$ и $q$ and check out which of them are stable near the minimum of the free energy $F = -k_B T \ln Z$. Due to the exchange interaction $J$ the antiferromagnetic (AFM) HS ground state exists up to $P = P_C > P_{C0}$ (Fig.2b). Increasing the critical pressure due to effect of cooperativity is expected because the exchange interaction stabilizes the HS state. At $P > P_C$ the ground LS- state takes place, while the crystal volume reveals a sharp decrease at $P = P_C$ (Fig.2c). The HS-LS sharp crossover at low temperatures and smooth one at high temperatures are seen in Fig.2a.

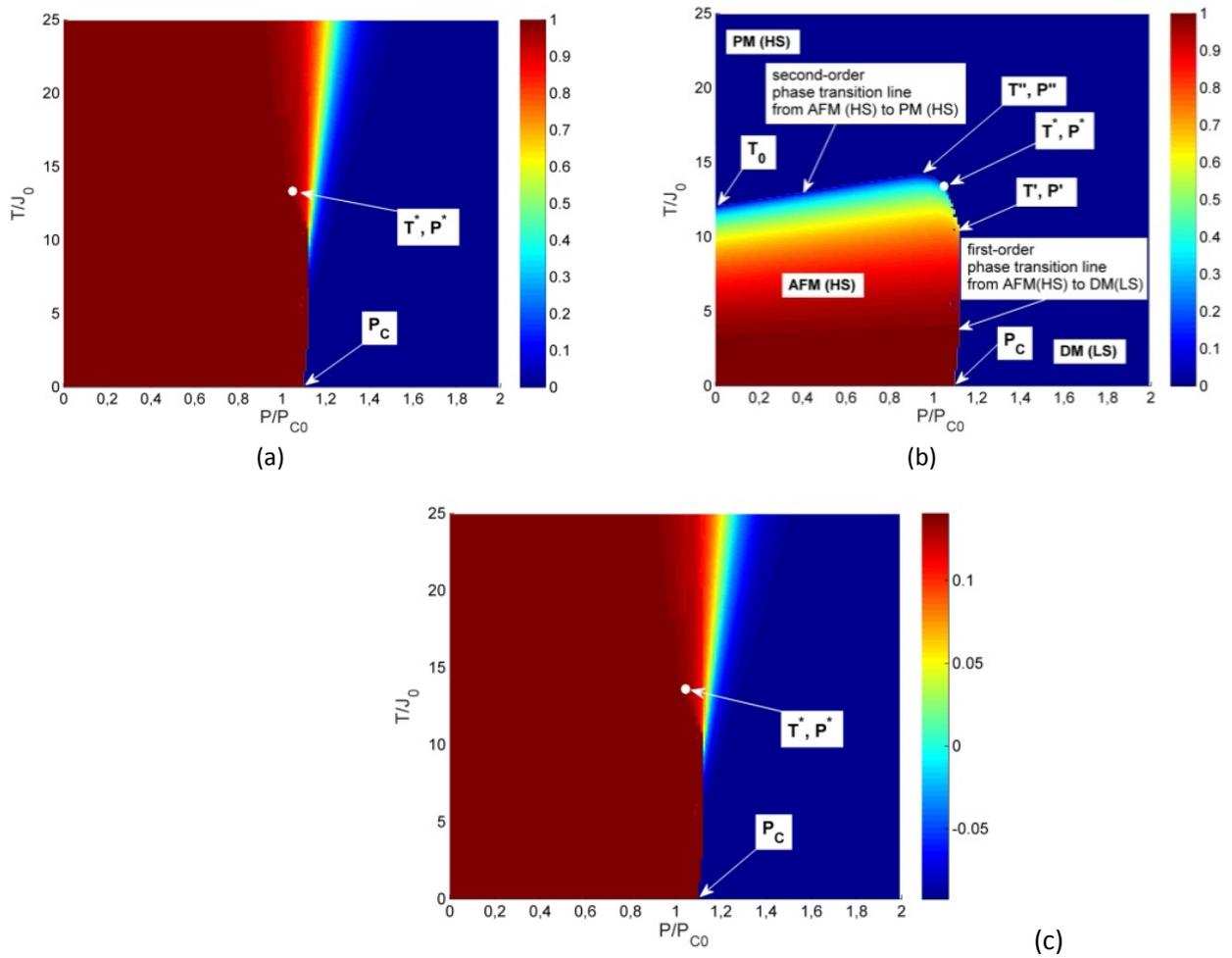

Fig. 2. The $P$-$T$ maps of the HS/LS concentration (a), magnetization (b) and lattice distortion (c) for the mixing value $J_x = 0.01$ eV.

In the HS area $P < P_C$ (Fig.2b) with increasing temperature we found the second order phase transition from the AFM to the paramagnetic phase when $P < P^*$ and the first order transition, when $P^* < P < P_C$. In the first case we see the smooth volume change, while in the second case the volume reveals the sharp change (Fig.2c). In the tricritical point ($T^*$ и $P^*$ in Fig.2b) the line of second order phase transitions smoothly transforms in the line of the first order phase transitions. In the interval $P_C < P \leq P'$ the ground state is non magnetic, but with heating the magnetic HS state is populated and the long range AFM HS-state appears (Fig.2b) with the sharp change of the volume (Fig.2c). Thus, due to cooperativity the reentrant magnetic transition in the vicinity of the crossover appears. With further heating the AFM–PM transition is of the second order if $P_C < P \leq P^*$ and the first order close to the second if $P^* < P \leq P'$. At $P > P'$ (Fig.2b) the non magnetic LS phase is stable for all temperatures. For these pressures with heating there is a smooth crossover from the non magnetic to diluted paramagnetic state.

Besides reentrant magnetization with increasing temperature at $P_C < P \leq P'$, we also have noticed reentrant behavior with increasing pressure for temperatures $T_0 < T \leq T'$, where $T_0$ is the Neel temperature for $P = 0$ and $T'$ is the maximal value of the Neel temperature, increasing due to pressure dependent exchange interaction. For $T_0 < T \leq T'$ the paramagnetic state at low pressure undergo second order transition to AFM and with forthcoming pressure increase transforms again into paramagnetic phase by the second order transition if $T^* < T_0$ or $T^* > T_0$, $T^* < T < T'$ and first order transition if $T^* > T_0$ and $T_0 < T < T^*$, (Fig.2b). For our set of parameters $T^* > T_0$. The volume with increasing pressure has a sharp drop if $0 \leq T \leq T^*$ or changes continuously if $T > T^*$, (Fig.2c).

Increasing of the mixing term $J_x$ has strong effect on the magnetization with reducing the Neel temperature with increasing pressure. At the same time the $n(P)$ and $q(P)$ dependences show smaller changes, mainly more smooth crossover. The reason of strong suppression of magnetization is clear from the structure of the Hamiltonian (4), where the operator $X^{s,\sigma}$ transforms the HS state with spin projection $\sigma$ into the LS singlet state $|s\rangle$. The phase diagram for $J_x = 0.05$ eV is shown with more details in Fig. 3.

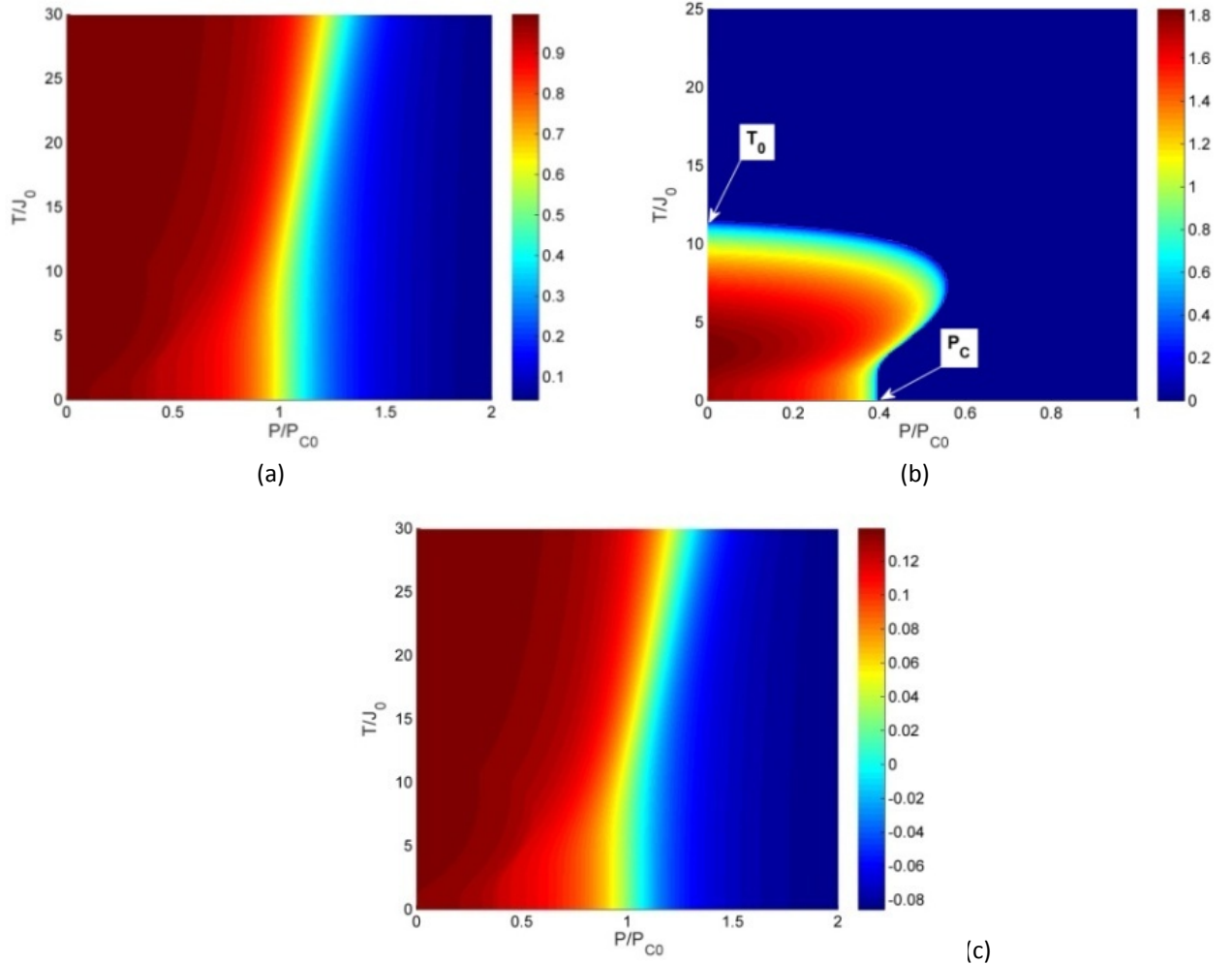

Fig. 3. The $P$-$T$ maps of the HS concentration $n$ (a), magnetization $m$ (b) and lattice distortion $q$ (c) for the mixing value $J_x = 0.05$ eV.

The Neel temperature at zero pressure $T_0/J_0 \sim 12$ is almost the same in Fig.2 and Fig.3, while the temperature dependence of the magnetization changes remarkable with increasing the value $J_x$. The existence of reentrant magnetic behavior with increasing temperature is clearly seen (Fig.3b), but in contrast to the previous case ($J_x = 0.01$ eV), the region of existence of long-range magnetic order decreases. So, there is a significant decrease of $P_C$ ($P_C < P_{C0}$) (Fig.3b). In addition, a significant difference are the smooth change of the magnetization $m$ at $T = 0$, the absence of the first order phase transitions and tricritical point in the phase diagram. As concerns the HS concentration $n$ and lattice distortion $q$, they are smoothly spread over pressure range shown by the color map in Fig. 3 a, c. All systems characteristics are changing continuously, by a second order phase transition.

## 4. Non-equilibrium quantum dynamics and relaxation processes

To study the relaxation processes, one should include the interaction with an environment. In what follows, we consider coupling of the spin crossover system with a common phonon environment. The total system–environment Hamiltonian can be written as

$$\hat{H} = \hat{H}_0 + \hat{H}_R + \hat{V}. \tag{14}$$

Here $\hat{H}_0 \equiv \hat{H}_{MF} = \sum_k E_k |\varphi_k\rangle\langle\varphi_k|$ is the Hamiltonian (5) of our spin crossover system written in the terms of its mean field eigenstates; $\hat{H}_R = \sum_q \hbar\omega_q b_q^\dagger b_q$ is the environment Hamiltonian with $b_q^\dagger(b_q)$ being phonon creation (annihilation) operators, and $\hat{V} = \hat{V}_{v-ph} + \hat{V}_{s-ph}$ describes the interaction of the environment with our spin crossover system, that includes the vibron – phonon $g_{v-ph}$ and spin-phonon $g_{s-ph}$ interactions given by

$$\hat{V}_{v-ph} = \sum_q \left( g_{v-ph,q} b_q^\dagger a + g^*_{v-ph,q} b_q a^\dagger \right),$$

$$\hat{V}_{s-ph} = \sum_q \left( g_{s-ph,q} b_q^\dagger \hat{S}^+ + g^*_{s-ph,q} b_q \hat{S}^- \right).$$

We use the reduced density-matrix approach, leading to the master equation:

$$\frac{d\hat{\rho}^0}{dt} = -i\left[\hat{H}_0(t), \hat{\rho}^0\right] + \hat{L}\rho^0, \tag{15}$$

where the superoperator $\hat{L}$ describes coupling to the environment. In what follows we assume that spin-crossover system is weakly coupled to phonon environment. This allows us to use the Markovian quantum master equation for the reduced density matrix. Further we assume also that the environment is short correlated, i.e. the environment will quickly "forget" its interactions with the system. Then Eq. (15) can be recast as the Redfield master equation [54-57]. The Redfield equation [58] for the reduced density matrix $\rho_{kl}^0$ may be written as

$$\frac{d}{dt}\rho_{kl}^0 = -i\omega_{kl}\rho_{kl}^0 - \sum_{m,n} \rho_{mn}^0 R_{klmn}. \tag{16}$$

The first term in Eq.(16) describes the reversible motion in terms of the transition frequencies $\omega_{kl} = \frac{E_k - E_l}{\hbar}$ between energy levels in the spin-crossover system, and the second term describes relaxation. The Redfield approximation is valid for time intervals $\Delta t \gg \tau_c$, where $\tau_c$ is the correlation time of the environment. The second simultaneous condition is [55] $R_{klmn}\Delta t \ll 1$. For the spin-crossover system with two channels of interaction with the environment, the relaxation matrix, $R_{klmn}$, reads $R_{klmn} = \sum_p \delta_{nl}\Gamma^+_{kppm} + \sum_p \delta_{km}\Gamma^-_{nppl} - \Gamma^+_{nlkm} - \Gamma^-_{nlkm}$.

Here the $\Gamma$ is determined by

$$\Gamma^+_{mkln} = \frac{1}{\hbar^2}\int_0^\infty dt \exp(-it\omega_{ln})Tr_R(V_{mk}(t)V_{ln}(0)\rho_R(0)), \quad (17)$$

$$\Gamma^-_{mkln} = \frac{1}{\hbar^2}\int_0^\infty dt \exp(-it\omega_{mk})Tr_R(V_{mk}(0)V_{ln}(t)\rho_R(0)).$$

$V_{mk}(t)$ are the matrix elements of the operator $\hat{V}$ in the interaction representation. In the secular approximation with $E_k - E_m + E_n - E_l = 0$ the Redfield equation (16) can be written as the generalized Master equation

$$\frac{\partial}{\partial t}\rho^0_{kl} = -i\omega_{kl}\rho^0_{kl} + \delta_{kl}\sum_{n \neq l}\rho^0_{nn}W_{ln} - \gamma_{kl}\rho^0_{kl}, \quad (18)$$

where $W_{ln} = \Gamma^+_{nlln} + \Gamma^-_{nlln}$, $\gamma_{kl} = \sum_n (\Gamma^+_{knnk} + \Gamma^-_{lnnl}) - \Gamma^+_{llkk} - \Gamma^-_{llkk}$. For the diagonal matrix elements it looks as

$$\frac{\partial}{\partial t}\rho^0_{kk}(t) = \sum_{n \neq k}\rho^0_{nn}(t)W_{kn} - \rho^0_{kk}(t)\sum_{n \neq k}W_{nk}. \quad (19)$$

For any dynamical operator $\hat{Q}_0$ (in our case these are operators $\hat{m}$, $\hat{n}$, $\hat{q}$), the mean value is equal to: $\langle \hat{Q}_0 \rangle = Tr\hat{Q}_0\hat{\rho}^0(t)$.

To calculate the relaxation tensor components we write down the interaction with environment as

$$\hat{V}_{\upsilon-ph} = \sum_q \left( g_{\upsilon-ph,q}b_q^\dagger \sum_{ij}|\varphi_i\rangle\langle\varphi_i|a|\varphi_j\rangle\langle\varphi_j| + g^*_{\upsilon-ph,q}b_q\sum_{ij}|\varphi_i\rangle\langle\varphi_i|a^\dagger|\varphi_j\rangle\langle\varphi_j|\right),$$

$$\hat{V}_{\upsilon-ph} = \sum_{ij,q}(g_{\upsilon-ph,q}b_q^\dagger a_{ij} + g^*_{\upsilon-ph,q}b_q a_{ji})|\varphi_i\rangle\langle\varphi_j| = \sum_{ij}\hat{B}^{\upsilon-ph}_{ij}\hat{Q}_{ij},$$

where $\hat{B}^{\upsilon-ph}_{ij} = \sum_q (g_{\upsilon-ph,q}b_q^\dagger a_{ij} + g^*_{\upsilon-ph,q}b_q a_{ji})$, $\langle\varphi_i|a|\varphi_j\rangle = a_{ij}$, and $\hat{Q}_{ij} = |\varphi_i\rangle\langle\varphi_j|$.

Similarly for the spin-phonon interaction $\hat{V}_{s-ph} = \sum_{ij}\hat{B}^{s-ph}_{ij}\hat{Q}_{ij}$, where $\hat{B}^{s-ph}_{ij} = \sum_q (g_{s-ph,q}b_q^\dagger s_{ij} + g^*_{s-ph,q}b_q s_{ji})$, $\langle\varphi_i|\hat{S}^+|\varphi_j\rangle = s_{ij}$.

Finally, for the total interaction with environment

$$\hat{V} = \sum_{ij}(\hat{B}^{\upsilon-ph}_{ij} + \hat{B}^{s-ph}_{ij})\hat{Q}_{ij} = \sum_{ij}\hat{B}_{ij}\hat{Q}_{ij}, \quad \hat{B} = (\hat{B}_{\upsilon-ph} + \hat{B}_{s-ph})$$

and the Eq.(17) can be written as

$$\Gamma^+_{mkln} = \int_0^\infty d\tau e^{-i\omega_{ln}\tau}\langle\hat{B}_{mk}\hat{B}_{ln}(\tau)\rangle_R, \quad \Gamma^-_{mkln} = \int_0^\infty d\tau e^{-i\omega_{mk}\tau}\langle\hat{B}_{mk}(\tau)\hat{B}_{ln}\rangle_R, \quad (20)$$

where $\hat{B}(\tau) = e^{i\hat{H}_{ph}\tau}\hat{B}e^{-i\hat{H}_{ph}\tau}$.

For the vibron-phonon relaxation channel the straightforward calculations results in

$$\Gamma_{nlln}^{(\upsilon-ph)+} = a_{nl}^2 \int_0^\infty \sum_q \left|g_{\upsilon-ph,q}\right|^2 \langle \hat{n}_q \rangle_R e^{i(\omega_q - \omega_{ln})\tau} d\tau + a_{ln}^2 \int_0^\infty \sum_q \left|g_{\upsilon-ph,q}\right|^2 \left(\langle \hat{n}_q \rangle_R + 1\right) e^{-i(\omega_q - \omega_{nl})\tau} d\tau,$$

that can be estimated as

$$\Gamma_{nlln}^{(\upsilon-ph)+} \approx \gamma_0^{\upsilon-ph} \left[ a_{nl}^2 n_{BE}(\omega_{ln}) + a_{ln}^2 \left(n_{BE}(\omega_{nl}) + 1\right) \right], \quad (21)$$

where $n_{BE}$ is the Bose-Einstein distribution function. Similar, for the spin-phonon relaxation channel

$$\Gamma_{nlln}^{(s-ph)+} \approx \gamma_0^{s-ph} \left[ s_{nl}^2 n_{BE}(\omega_{ln}) + s_{ln}^2 \left(n_{BE}(\omega_{nl}) + 1\right) \right]. \quad (22)$$

The product $V_{\upsilon-ph}(t) V_{s-ph}(0)$ in Eq.(17) gives zero contribution due to the matrix elements. Finally,

$$\Gamma_{nlln}^+ = \Gamma_{nlln}^{(\upsilon-ph)+} + \Gamma_{nlln}^{(s-ph)+}. \quad (23)$$

The averaged over the Brillouin zone parameters of the vibron-phonon and spin-phonon coupling are two external parameters that characterize the dynamics of our system. To minimize the number of free parameters we take them to be equal. As a typical value we accept these parameters to be $\tau_0^{\upsilon-ph} \sim \frac{1}{\gamma_0^{\upsilon-ph}} \sim 1\,\text{ps}$, $\tau_0^{s-ph} \sim \frac{1}{\gamma_0^{s-ph}} \sim 1\,\text{ps}$. We want to emphasize that in spite of equal values of the parameters $\gamma_0$ for the vibron-phonon and spin-phonon interactions the magnetic and lattice relaxations times would be different due to the different matrix elements in Eqs.(21), (22).

We have calculated coefficients $\Gamma$ in Eqs.(21-23) and its linear combinations $W_{ln}$ and $\gamma_{kl}$ enter the generalized Master equation. Some information on these coefficients is given in Fig. 4.

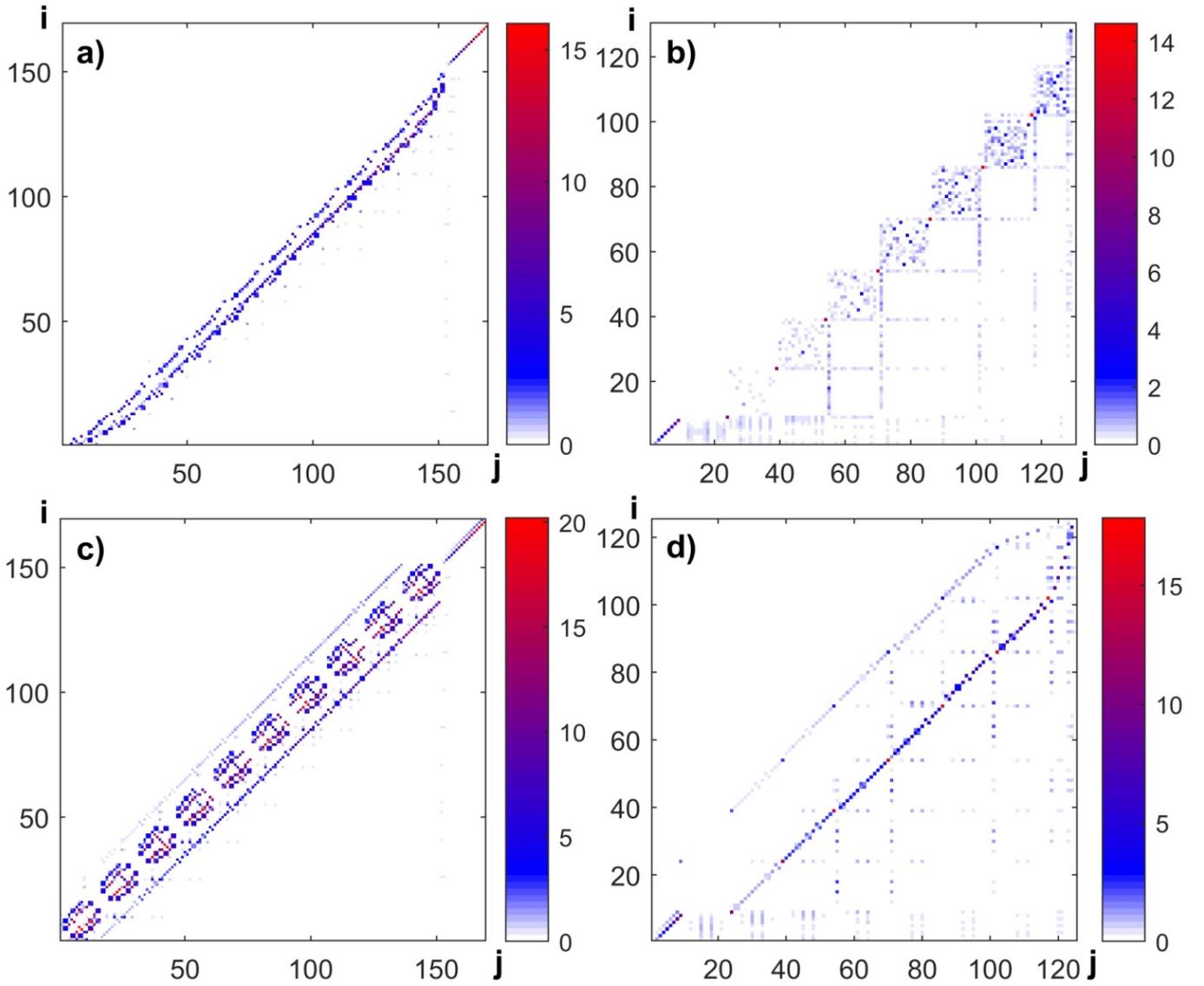

Fig. 4. The structure of relaxation matrix elements $\Gamma$ from the eqn.(23) is shown for a set of 169 low energy level for $P/P_{C0} = 0.1$ (HS ground state) for $T = 0$ (a) and 300K (c) and up to 130 energy levels for $P/P_{C0} = 1.5$ (LS ground state) for $T = 0$ (b) and 300K (d). The spin-orbital parameter $J_x = 50$ meV.

For both pressures we can see in Fig. 4 the formation of a system of sublevel clusters containing 16 levels. The origin of these clusters is clear from the discussion of the eigenstate (7) structure. At zero temperature we can see only under diagonal matrix elements corresponding to the excitations from occupied into non occupied states. At $T = 300$ K the matrix is almost symmetrical indicating up and down excitations.

### 5. Numerical results for system dynamics at various pressure

We assume that ground at $P < P_C$ HS state may be suddenly excited (for example, by ultrashort ligh pulse) into the LS state, or at $P > P_C$ the ground LS state excited in the HS state. In the experiment [11] such excitations consists of two steps. Initially the $d - p$ excitation occurs under light pulse from the LS term into some higher energy states, and then the relaxation to the LS state takes place. The total time for the excitation between two different spin states takes the period of time of the order 10 - 100 fs. In our simulations we assume that the ground state for the given pressure $|\varphi_0\rangle = \sum_{n_{ph}=0}^{N_{ph}} \left[ a_{n_{ph},0} |2,0,n_{ph}\rangle + \sum_{s_z=-S}^{+S} b_{n_{ph},s_z,0} |1,s_z,n_{ph}\rangle \right]$ can be suddenly excited at $t=0$ into the non-equilibrium initial state $|\psi_0\rangle$. Its structure depends on pressure. For $P < P_C$ and $T = 0$ when coefficients in the ground state $|\varphi_0\rangle$ $b$ is close to 1 and $a$ is close to zero, we write down the initial state as $|\psi_0\rangle = \sum_{n_{ph}=0}^{N_{ph}} \left[ a_{n_{ph},0} |2,0,n_{ph}\rangle + \sum_{s_z=-S}^{+S} b_{n_{ph},s_z,0} |2,0,n_{ph}\rangle \right]$ by switching HS→LS, while the lattice is not excited and remains in the initial state with $q_{HS}^0$. And for $P > P_C$ vice versa, the ground state is the LS one with coefficient $a \sim 1$, $b \sim 0$. So we write down at $t=0$ the excited initial state to be in the HS state. It is given by $|\psi_0\rangle = \sum_{n_{ph}=0}^{N_{ph}} \left[ a_{n_{ph},0} |1,+2,n_{ph}\rangle + \sum_{s_z=-S}^{+S} b_{n_{ph},s_z,0} |1,+2,n_{ph}\rangle \right]$ and the lattice is in the LS state.

The excited state can be written in the eigenstate basis (7) as $|\psi_0\rangle = \sum_k C_{0k} |\varphi_k\rangle$ with $C_{0k} = \langle \varphi_k | \psi_0 \rangle$. The initial density matrix is equal to $\rho_{kk'}^0(0) = C_{0k} C_{0k'}^*$. For finite temperatures

$$\hat{\rho}^0(0) = \sum_k \frac{\exp\left(-\frac{E_k}{k_B T}\right)}{Z} |\psi_k\rangle\langle\psi_k| = \sum_k \sum_{ii'} \frac{\exp\left(-\frac{E_k}{k_B T}\right)}{Z} C_{ik} C_{i'k}^* |\varphi_i\rangle\langle\varphi_{i'}|, \quad \text{where} |\psi_k\rangle = \sum_i C_{ki} |\varphi_i\rangle,$$
$C_{ki} = \langle \varphi_i | \psi_k \rangle$.

The relaxation dynamics of magnetization $m$ (red line), HS occupation number $n$ (blue line) and lattice distortion $q$ (black line) is shown for two values of the mixing energy $J_x = 0.01$ eV (Fig. 5) and $J_x = 0.05$ eV (Fig. 7). We take into account 3-fold orbital degeneracy of the HS state. The temperature was fixed $T = 100$ K, while the pressure was varied from $0.1 P_{C0}$ to $1.5 P_{C0}$.

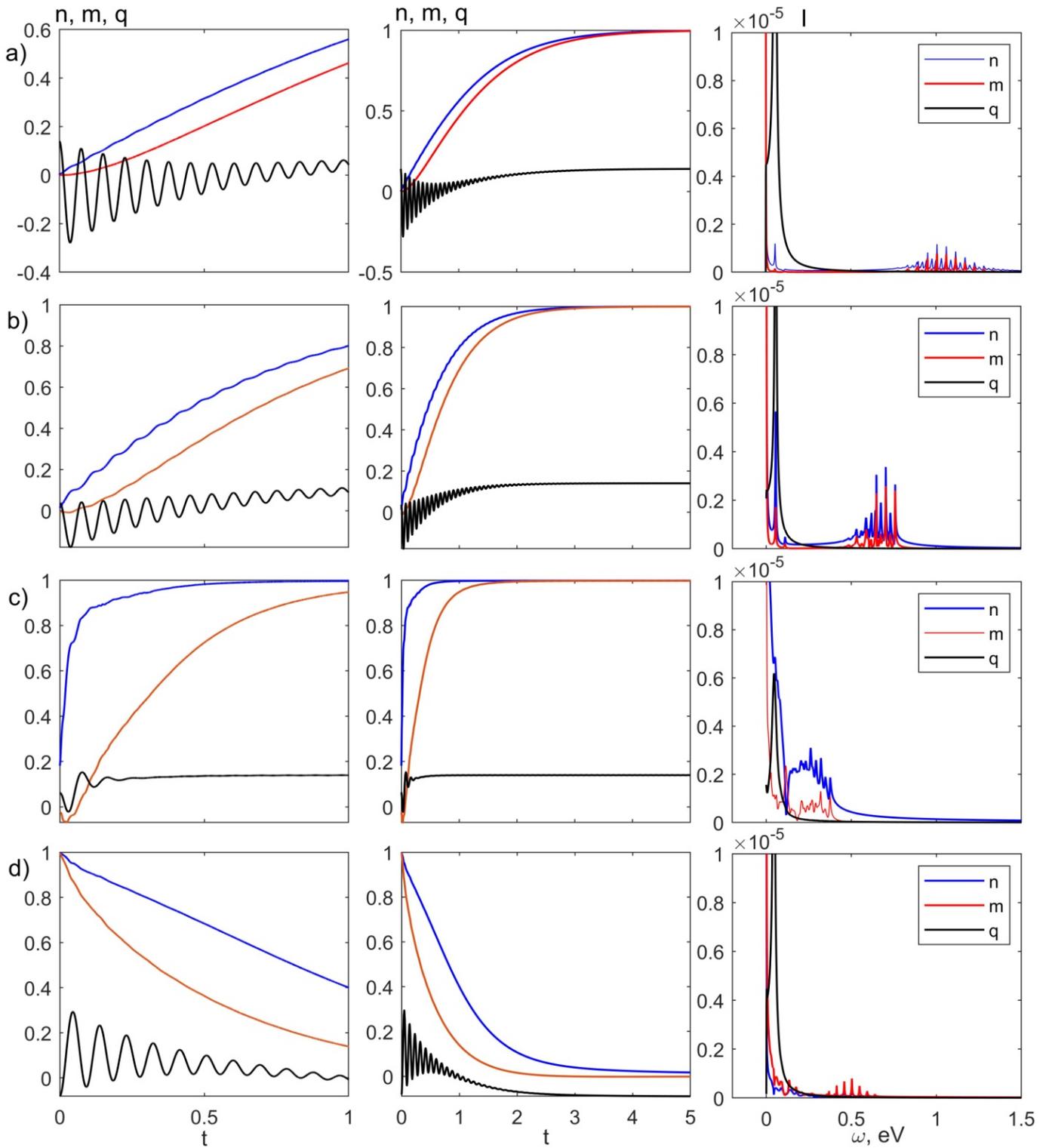

Fig. 5. Quantum dynamics of the photoexcited Franck-Condon states relaxation in magnetic insulator with spin crossover at $T = 100$ K and pressure $P/P_{C0} = 0.1$ (a), $P/P_{C0} = 0.5$ (b), $P/P_{C0} = 1$ (c), and $P/P_{C0} = 1.5$ (d) for $J_x = 0.01$ eV. In the first column we show initial stage of relaxation with the time from 0 to 1 in the units $\tau_0 = 10^{-12}$ sec, in the second column the same up to $t = 5$. The right column shows the Fourier transforms of $m$ (red line), HS- concentration $n$ (blue line), and lattice distortion $q$ (black line).

To find the relaxation times we fit the data from Fig. 5 and Fig. 7 by the exponential law $y_i = y_{0i} + \eta_i e^{-\xi_i t}$, $i = m, n, q$ and $\eta_i$ и $\xi_i$ are fitted parameters while the equilibrium value $y_{0i}$ was taken from the mean field phase diagrams. Fig. 6 and Fig. 9 show the results of such fitting for the low pressure $P/P_{C0} = 0.1$ (upper line) and high pressure $P/P_{C0} = 1.5$ (low line) at $J_x = 0.01$ eV and $J_x = 0.05$ eV, correspondingly.

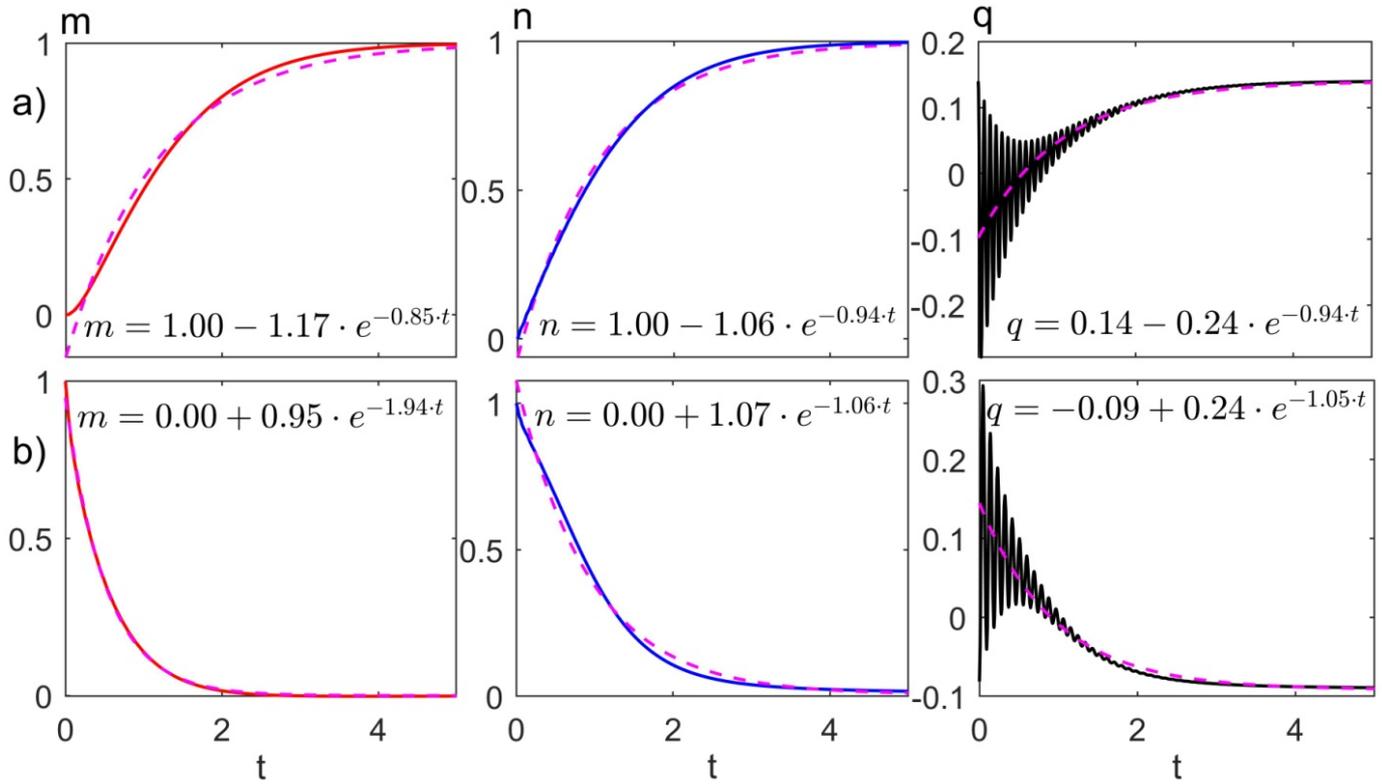

Fig. 6. Exponential fitting $y = y_0 + \eta e^{-\xi t}$ of dynamics $m$, $n$, and $q$ for $P/P_{C0} = 0.1$ (above) and $P/P_{C0} = 1.5$ (down) at $J_x = 0.01$ eV.

Comparison of Fig. 5 and Fig. 6 has revealed that the non equilibrium magnetization $m$, HS- concentration $n$, and lattice distortion $q$ tend to its equilibrium values with different relaxation time. For magnetization the time is $t_m$, while relaxation times for $n$ and $q$ practically equal. This agreement is not occasional because the change of the bond length $q$ is proportional to the cation radii. The analysis of relaxation times is given in Table 1.

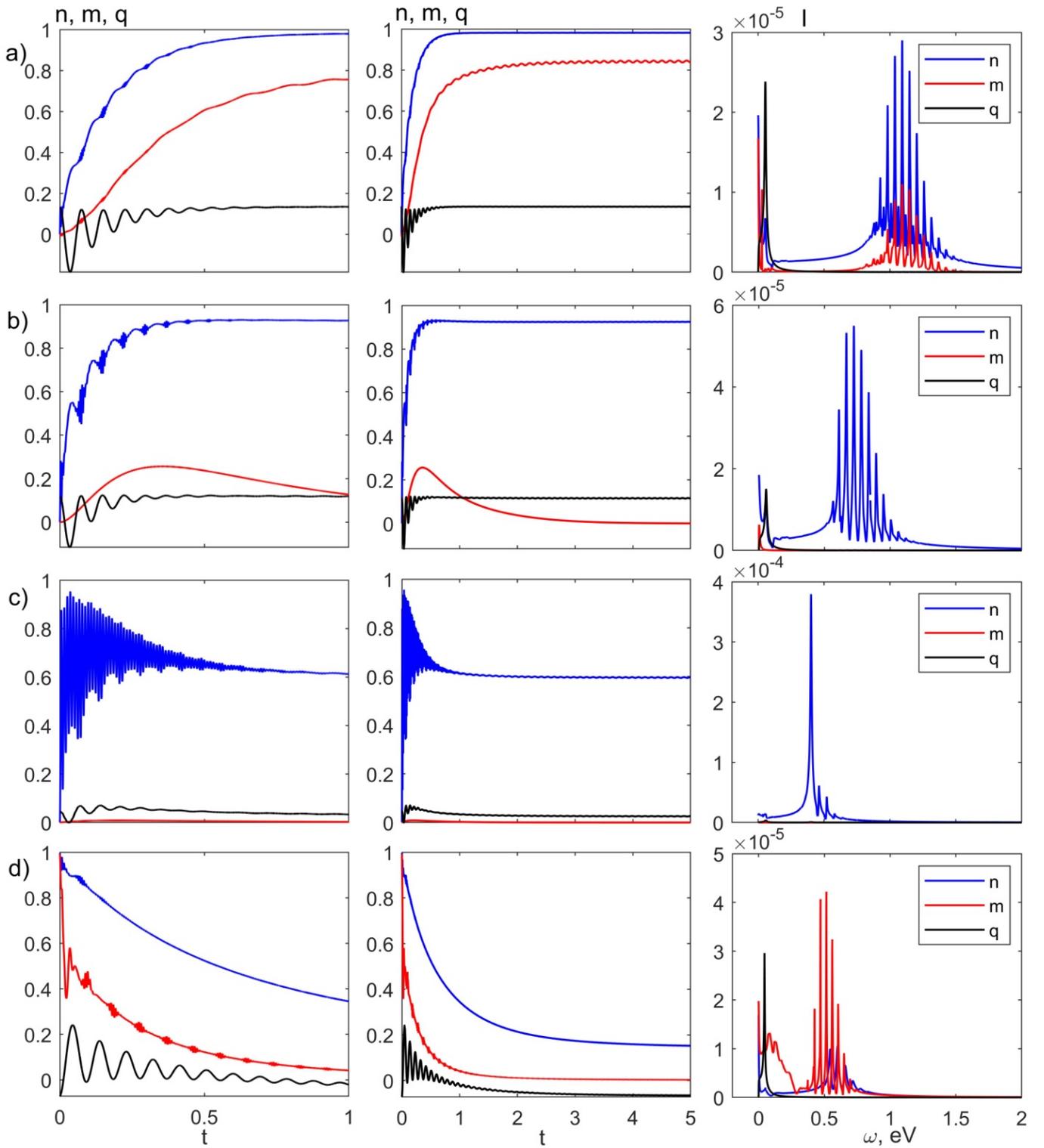

Fig. 7. Quantum dynamics of the photoexcited Franck-Condon states relaxation in magnetic insulator with spin crossover at $T = 100$ K and pressure $P/P_{C0} = 0.1$ (a), $P/P_{C0} = 0.5$ (b), $P/P_{C0} = 1$ (c), and $P/P_{C0} = 1.5$ (d) for $J_x = 0.05$ eV. In the first column we show initial stage of relaxation with the time from 0 to 1 in the units $\tau_0 = 10^{-12}$ sec, in the second column the same up to $t = 5$. The right column shows the Fourier transforms of $m$ (red line), HS- concentration $n$ (blue line), and lattice distortion $q$ (black line).

The strong change of magnetic dynamics in Fig. 7b,c vs Fig. 5b,c is related to the strong suppression of the critical pressure in Fig.4b. For $P/P_{C0} = 0.5$ $m(T=0)=0$, but the reentrant magnetization for this pressure appears for the temperature interval $4 < T/J_0 < 10$. This reentrance magnetization appears also dynamically with maximal value at $t = 0.25$ in Fig.7b. Nevertheless the equilibrium state is nonmagnetic in agreement to the phase diagram. In Fig.7c we see the large amplitude for initial oscillations of the HS- concentration $n$. In the phase diagram for $P/P_{C0} = 1$ and $T = 0$ the value of $n \sim 0.8$ with a smooth distribution from $n = 1$ at $P/P_{C0} = 0.8$ till $n = 0$ for $P/P_{C0} = 1.2$. This wide distribution reveals itself also in temporal scale.

Comparison $P/P_{C0} = 0.1$ (a) and $P/P_{C0} = 1.5$ (d) in Fig. 7 shows that for HS $n$ и $q$ has faster relaxation then $m$, that demonstrates the long precession. Contrary, for LS state $m$ and $q$ has faster relaxation, while attenuation of $n$ is slower. For $t = 5$ $n$ is still quite large, $n = 0.2$ instead of expected zero value. It has been shown before that for some dynamical regime of loading the stationary state may be a mixture of the HS- and LS-states [59], that is why we have specially checked the regime $P/P_{C0} > 1$ up to 80 ps and found that suddenly excited HS- state for $T = 300$ K gradually relaxes to the stationary LS- state (Fig. 8).

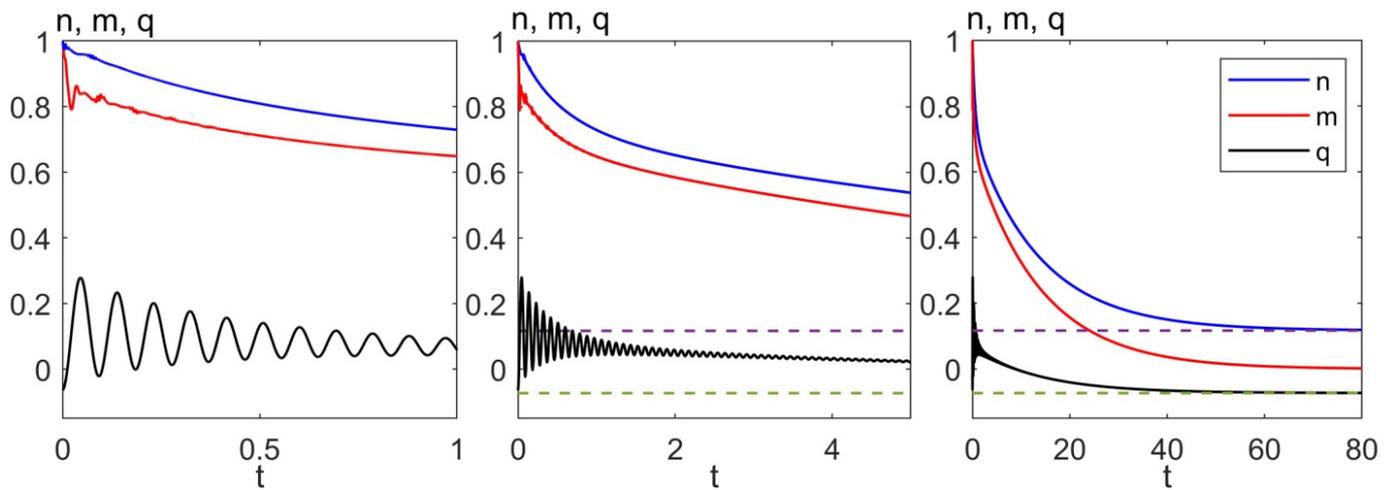

Fig. 8. Slow relaxation of the suddenly excited HS- state at $P/P_{C0} = 1.5$, $J_x = 0.05$ eV and $T = 300$ K to the stationary LS- state shown by dotted lines for $n$ and $q$. For magnetization the stationary state is $m = 0$.

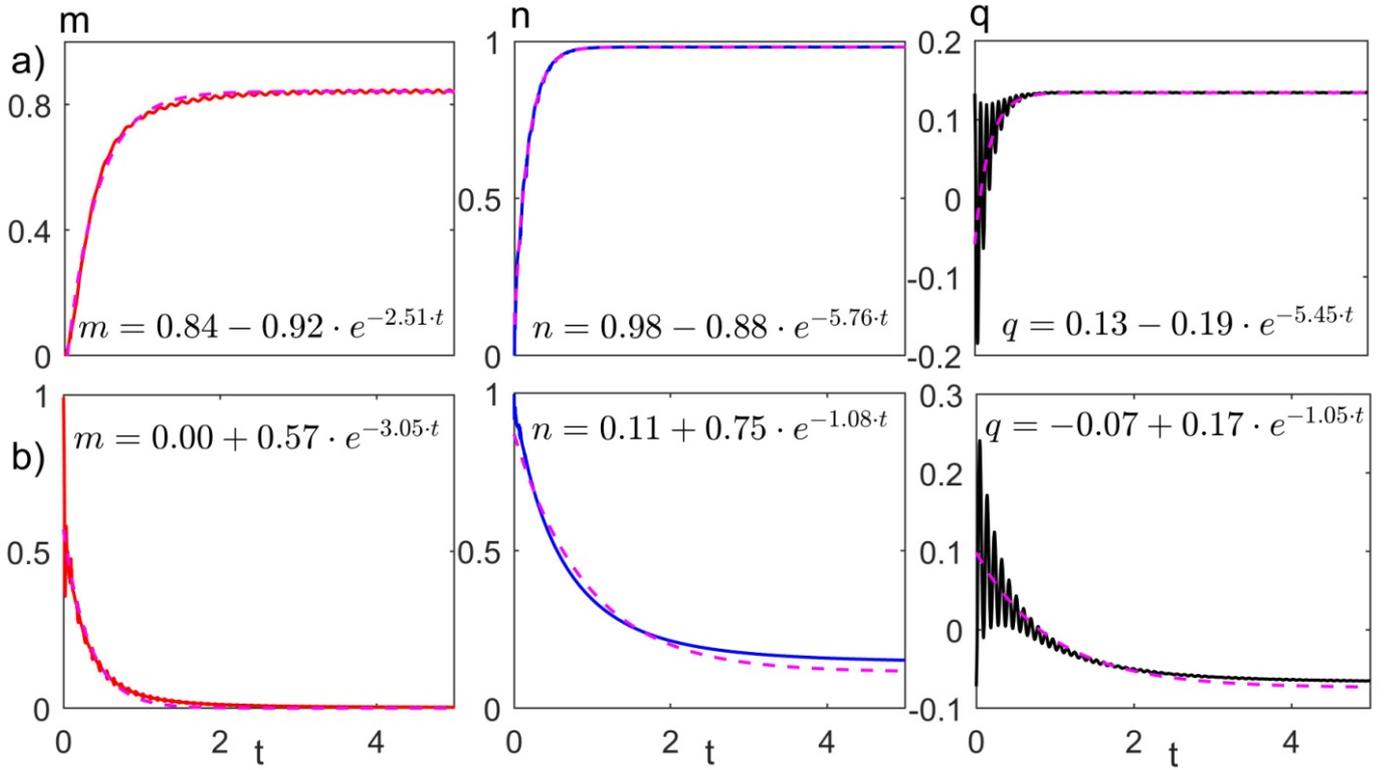

Fig. 9. Exponential fitting $y = y_0 + \eta e^{-\xi t}$ of dynamics $m$, $n$, and $q$ for $P/P_{C0} = 0.1$ (above) and $P/P_{C0} = 1.5$ (down) at $J_x = 0.05$ eV.

| $J_x$, meV | $P/P_{C0}$ | $\tau_m$, ps | $\tau_n$, ps | $\tau_q$, ps |
|---|---|---|---|---|
| 10 | 0.1 | 1.18 | 1.06 | 1.06 |
|  | 1.5 | 0.52 | 0.94 | 0.95 |
| 50 | 0.1 | 0.40 | 0.18 | 0.18 |
|  | 1.5 | 0.33 | 0.93 | 0.95 |

Table 1. Relaxation times for magnetization ($\tau_m$), HS concentration ($\tau_n$) and bond length ($\tau_q$) for different pressures and mixing interaction $J_x$.

**VI. Discussion of results**

We summarize the values of the relaxation times and oscillation frequencies from the Figs. 5, 7 in Tab. 1,2. We note from Tab. 1 that all relaxation times decreases with larger spin-orbital term $J_x$ and this is absolutely evident. We obtain also opposite ratio for magnetic/nonmagnetic relaxation times at small and high pressures. Indeed, when $P/P_{C0} = 0.1$ and the HS state is the equilibrium one, the magnetic relaxation time $\tau_m$ is larger than the HS concentration relaxation time $\tau_n$ and lattice one $\tau_q$, for both spin-orbit coupling values. Contrary, when the ground state is the non magnetic LS at $P/P_{C0} = 1.5$, the magnetic relaxation is faster than the HS concentration and lattice relaxations. From Tab. 2 it is evident that the vibration frequencies are almost independent of the pressure and spin orbital interaction. As concerns the magnetization and the HS concentration oscillations that show a multimodal behavior, they have high frequency components for some pressure values beside the vibration frequency.

Spectral analysis in Figs. 5 and 7 reveals several time scales in the complex dynamics of the system. We found a remarkable difference of the dynamic for weak and strong spin-orbital interaction. Thus for $J_x = 0.05$ eV we can see in the $m(t)$ and $n(t)$ curves several regular and strongly non linear excitations like the short wave packets with oscillation energy $\sim 1$eV. Narrow peaks in the spectrum of this short packets in Fig. 7 are splitted with energy interval $\Delta\omega = 58$ meV, that agrees with the vibron energy $\omega_{LS} = 55$ meV. It allows us to relate these high frequency excitations with the Franck-Condon resonances that correlates with minima and maxima of $q(t)$ oscillations. These perturbations have the evolution with relaxation time $\sim \tau_q$. For small pressure $P/P_C = 0.1$ we can see long living periodic magnetic oscillations with period 140 fs and energy 35 meV. Similar magnetic oscillations are seen also for $J_x = 0.01$ eV in the Fourier spectra in Fig. 5 with smaller amplitude vs Fig. 7. Similar low frequency magnetic oscillations has been found at the femtosecond pumping of the weak ferromagnet FeBO$_3$ at normal pressure [60, 61]. In these experiments the initial HS ($S = 5/2$) state has been excited into the intermediate state of the Fe$^{3+}$ with spin $S = 3/2$, in 4 ps after excitation periodic oscillations of magnetization appeared with 2 ps period. In our calculations periodic magnetic oscillations with period 0.14 ps appeared after fast (2ps) relaxation of electronic and elastic systems to the equilibrium for the HS values. Our model has been developed for the $3d^6$ Fe$^{2+}$ oxides with the other HS and LS values then take place for Fe$^{3+}$ tem in FeBO$_3$. Moreover, it has too many arbitrary parameters то pretend for some qualitative agreement with experiment. Nevertheless the qualitative picture of the magnetic oscillations found in experiments [60, 61] and found in our calculations is similar.

| $J_x$, meV | $P/P_C$ | $\omega_m$, meV | $\omega_n$, meV | $\omega_q$, meV |
|---|---|---|---|---|
| 10 | 0.1 | FCR 1100 with $\Delta\omega = 55$ | 55 weak, FCR 1100 with $\Delta\omega = 27.5$ | 55 strong |
|  | 0.5 | 55 weak, FCR 675 with $\Delta\omega = 55$ | 55 strong, FCR 675 with $\Delta\omega = 27.5$ | 55 strong |
|  | 1.0 | 250 +/- 100 | 250 +/- 100 | 47 |
|  | 1.5 | FCR 500 with $\Delta\omega = 45$ | - | 45 |
| 50 | 0.1 | 29, FCR 1100 with $\Delta\omega = 55$ | 55 weak, FCR 1100 with $\Delta\omega = 27.5$ | 55 strong |
|  | 0.5 | - | 55, FCR 723 with $\Delta\omega = 57$ | 55 |
|  | 1.0 | - | 56, FCR 400 with $\Delta\omega = 60$ | 56 weak |
|  | 1.5 | wide FCR 84 with $\Delta\omega = 48$ FCR 515 with $\Delta\omega = 45$ | 45 weak, FCR 600 with $\Delta\omega = 45$ | 45 |

Table 2. Oscillation frequencies for magnetization, HS concentration and bond length for different pressures and spin orbital interaction $J_x$. Weak means small amplitude $\Delta\omega = 55$. FCR 675 is a set narrow equidistant Frank-Condon resonances centered at 675 meV.

We notice that at low pressure $P/P_{C0} = 0.1$ and $P/P_{C0} = 0.5$ with the HS ground state, after the sharp excitation of the electronic and magnetic systems in the LS state without changing the surrounding anions, the relaxation of the bond length is characterized by the frequency 55meV, corresponding to the LS oscillation frequency $\omega_{LS}$. And vice versa, for pressure $P/P_{C0}=1.5$ when the electronic and magnetic systems of the LS ground state is sharply excited in the HS initial state the relaxation is characterized by the HS frequency $\omega_{HS}=45$ meV. This fact demonstrates that the electronic, magnetic and elastic systems in SC materials are so strongly correlated that the fluctuation in one of them results in similar fluctuations of the others.

### VII. Conclusion

In magnetic materials with spin crossover the switching between HS and LS states is strongly related to the lattice degrees of freedom, that together with the interatomic exchange interaction provides the effects of cooperativity. Up to now the most part of experimental research of the ultrafast spin crossover dynamics have been carried out with non magnetic materials. In this paper we have found magnetization oscillations and complex multiscale dynamics of magnetic, HS concentration and Me-O bond length relaxation in strongly correlated electronic system with long range magnetic order. We hope that our theory may stimulate more experimental research of the ultrafast magnetic dynamics.

**Acknowledgements**
The authors thank the Russian Scientific Foundation for the financial support under the grant 18-12-00022.